\documentclass[11pt, a4paper]{article}

\ifcsname dgB@loaded\endcsname\expandafter\endinput\fi
\expandafter\let\csname dgB@loaded\endcsname\empty

\usepackage{ifthen}
\usepackage{etoolbox}%

\makeatletter
\@ifundefined{ifdelim_grow}{%
\newboolean{delim_grow}\setboolean{delim_grow}{true}%
}{}
\@ifundefined{ifare_slides}{%
\newboolean{are_slides}%
\setboolean{are_slides}{false}%
}{}
\@ifundefined{ifdg_impose_style}{%
\newboolean{dg_impose_style}\setboolean{dg_impose_style}{true}%
}{}
\@ifundefined{ifdg_load_tikz}{%
\newboolean{dg_load_tikz}\setboolean{dg_load_tikz}{true}%
}{}
\makeatother

\usepackage{amssymb}
\usepackage{amsmath}%

\usepackage{amsthm}

\usepackage{latexsym}

\ifthenelse{\boolean{are_slides}}{}{\usepackage{enumitem}}

\usepackage{units}%

\usepackage{filemod}%

\usepackage[nolabel]{fnbreak}%

\usepackage{psfrag}%

\makeatletter
\@ifpackageloaded{xcolor}{%
\@ifundefinedcolor{BrickRed}{\input{dvipsnam.def}}{}%
}{\usepackage[dvipsnames]{xcolor}}
\makeatother
\ifthenelse{\boolean{dg_load_tikz}}{\usepackage{tikz}}{}

\makeatletter
\@ifpackageloaded{url}{}{\PassOptionsToPackage{hyphens,spaces,obeyspaces}{url}}
\makeatother
\usepackage{url}%

\usepackage{graphicx}

\usepackage{eso-pic}%

\usepackage{marginnote}%

\usepackage{caption}%

\usepackage{calc}%

\usepackage{soul}%
\setul{1.25pt}{.16pt}%

\usepackage{bm}%

\usepackage{newunicodechar}%
\newunicodechar{⇒}{\ensuremath{\Rightarrow}}
\newunicodechar{ꞏ}{\scalebox{0.75}{\ensuremath{\bullet}}}

\ifthenelse{\boolean{dg_impose_style}}{%
\usepackage{tempora}%

\usepackage[T1,OT2,T2A]{fontenc}

\usepackage[utf8]{inputenc}

\usepackage[shorthands=off,greek,czech,ukrainian,english]{babel}

\DeclareTextSymbolDefault{\CYRYAT}{OT2}
\DeclareTextSymbolDefault{\cyryat}{OT2}
\DeclareTextSymbolDefault{\CYRFITA}{OT2}
\DeclareTextSymbolDefault{\cyrfita}{OT2}
\DeclareTextSymbolDefault{\CYRIZH}{OT2}
\DeclareTextSymbolDefault{\cyrizh}{OT2}

\fontencoding{OT1}\selectfont%
\usepackage[onlymath]{sansmathfonts}
\fontencoding{T2A}\selectfont

}{}%

\usepackage{microtype}%

\usepackage{mathrsfs}%

\usepackage{calligra}%

\usepackage{pdfrender}%

\usepackage{hyperref}%
\setpdflinkmargin{0.75pt}%
\ifthenelse{\boolean{dg_impose_style}}{%
\hypersetup{
colorlinks,
linkcolor={red!55!black},
citecolor={blue!55!black},
urlcolor={blue!35!black}
}
}{}

\ifthenelse{\boolean{are_slides}}{
\usepackage{beamerthemesplit}

\usepackage{ragged2e}%

\usetheme[secheader]{Boadilla}

\hypersetup{hidelinks}%

\usenavigationsymbolstemplate{}%
}{}%

\ifthenelse{\boolean{are_slides}}{
\newcommand{\MyChangeMathMargins}{%
\setlength{\abovedisplayskip}{\abovedisplayshortskip + 3pt}%
\setlength{\belowdisplayskip}{\abovedisplayshortskip + 4pt}%
}

\setul{1.5pt}{0.65pt}%
}{
\newcommand{\MyChangeMathMargins}{%
\setlength{\abovedisplayskip}{\abovedisplayshortskip + 4pt}%
\setlength{\belowdisplayskip}{\abovedisplayshortskip + 5pt}%
}

\ifthenelse{\boolean{dg_impose_style}}{%
\setlength\textheight{9.5in}

\setlength\topmargin{\dimexpr(\paperheight-\textheight
-\headheight-\headsep-\footskip)/2 - 1in\relax}

\setlength\textwidth{6.2in}

\setlength\oddsidemargin{\dimexpr(\paperwidth-\textwidth)/2 - 1in\relax}
\setlength\evensidemargin{\oddsidemargin}

\pdfpageheight=\paperheight
\pdfpagewidth=\paperwidth
}{}%
}%

\makeatletter

\newcommand{\NewcommandThreeArgsTwoOpt}[6][*]{%
\ifthenelse{\equal{}{#1}}%
{\let\comm@TMP\DeclareDocumentCommand}%
{\let\comm@TMP\NewDocumentCommand}%
\expandafter\def\csname\expandafter\@gobble\string#2@nopresq@nopostsq\endcsname##1##2{#3}%
\expandafter\def\csname\expandafter\@gobble\string#2@presq@nopostsq\endcsname##1##2{#4}%
\expandafter\def\csname\expandafter\@gobble\string#2@nopresq@postsq\endcsname##1##2##3{#5}%
\expandafter\def\csname\expandafter\@gobble\string#2@presq@postsq\endcsname##1##2##3{#6}%
\comm@TMP{#2}{o m o}{%
\IfNoValueTF{##1}{%
\IfNoValueTF{##3}%
{\csname\expandafter\@gobble\string#2@nopresq@nopostsq\endcsname{}{##2}}%
{\csname\expandafter\@gobble\string#2@nopresq@postsq\endcsname{}{##2}{##3}}%
}{%
\IfNoValueTF{##3}%
{\csname\expandafter\@gobble\string#2@presq@nopostsq\endcsname{##1}{##2}}%
{\csname\expandafter\@gobble\string#2@presq@postsq\endcsname{##1}{##2}{##3}}%
}%
}%
}

\newcommand{\NewcommandTwoArgsLastOpt}[4][*]{%
\ifthenelse{\equal{}{#1}}%
{\let\comm@TMP\DeclareDocumentCommand}%
{\let\comm@TMP\NewDocumentCommand}%
\expandafter\def\csname\expandafter\@gobble\string#2@nopostsq\endcsname##1{#3}%
\expandafter\def\csname\expandafter\@gobble\string#2@postsq\endcsname##1##2{#4}%
\comm@TMP{#2}{m o}{%
\IfNoValueTF{##2}%
{\csname\expandafter\@gobble\string#2@nopostsq\endcsname{##1}}%
{\csname\expandafter\@gobble\string#2@postsq\endcsname{##1}{##2}}%
}%
}

\newcommand{\NewcommandFourArgsTwoOpt}[6][*]{%
\ifthenelse{\equal{}{#1}}%
{\let\comm@TMP\DeclareDocumentCommand}%
{\let\comm@TMP\NewDocumentCommand}%
\expandafter\def\csname\expandafter\@gobble\string#2@nopresq@nopostsq\endcsname##1##2##3{#3}%
\expandafter\def\csname\expandafter\@gobble\string#2@presq@nopostsq\endcsname##1##2##3{#4}%
\expandafter\def\csname\expandafter\@gobble\string#2@nopresq@postsq\endcsname##1##2##3##4{#5}%
\expandafter\def\csname\expandafter\@gobble\string#2@presq@postsq\endcsname##1##2##3##4{#6}%
\comm@TMP{#2}{o m m o}{%
\IfNoValueTF{##1}{%
\IfNoValueTF{##4}%
{\csname\expandafter\@gobble\string#2@nopresq@nopostsq\endcsname{}{##2}{##3}}%
{\csname\expandafter\@gobble\string#2@nopresq@postsq\endcsname{}{##2}{##3}{##4}}%
}{%
\IfNoValueTF{##4}%
{\csname\expandafter\@gobble\string#2@presq@nopostsq\endcsname{##1}{##2}{##3}}%
{\csname\expandafter\@gobble\string#2@presq@postsq\endcsname{##1}{##2}{##3}{##4}}%
}%
}%
}

\newcommand{\NewcommandFourArgsTwoOptDefa}[5][*]{%
\ifthenelse{\equal{}{#1}}%
{\let\comm@TMP\DeclareDocumentCommand}%
{\let\comm@TMP\NewDocumentCommand}%
\comm@TMP{#2}{O{#3} m m O{#4}}{#5}%
}

\newcommand{\ifndef}[2]{\@ifundefined{#1}{#2}{}}

\newcommand{\IfMathMode}[2]{\ifmmode{#1}\else{#2}\fi}

\DeclareRobustCommand\bk{\@ifnextchar[%
{\lf\langle \bk@bra{}}%
{\@ifnextchar<%
{\lf.\bk@op}
{\lf| \bk@ket{}}%
}%
}

\def\bk@bra#1[#2]{\def\bk@tmp{#2}\ifx\bk@tmp\empty{#1}\else{#2}\fi\@ifnextchar[%
{\md|\hspace{-1.5pt}\md\langle \bk@bra{\def\bk@tmp{#2}\ifx\bk@tmp\empty{#1}\else{#2}\fi}}%
{\@ifnextchar\bgroup%
{\md| \bk@ket{\def\bk@tmp{#2}\ifx\bk@tmp\empty{#1}\else{#2}\fi}}%
{\@ifnextchar<%
{\md| \bk@op}%
{\rt|}%
}%
}%
}

\def\bk@ket#1#2{\def\bk@tmp{#2}\ifx\bk@tmp\empty{#1}\else{#2}\fi\@ifnextchar[%
{\md\rangle\hspace{-1.5pt}\md\langle \bk@bra{\def\bk@tmp{#2}\ifx\bk@tmp\empty{#1}\else{#2}\fi}}%
{\@ifnextchar\bgroup%
{\md\rangle\hspace{-1.5pt}\md| \bk@ket{\def\bk@tmp{#2}\ifx\bk@tmp\empty{#1}\else{#2}\fi}}%
{\@ifnextchar<%
{\md\rangle \bk@op}%
{\rt\rangle}%
}%
}%
}

\def\bk@op<#1>{#1 \@ifnextchar[%
{\md\langle \bk@bra{}}%
{\@ifnextchar\bgroup%
{\md| \bk@ket{}}%
{\@ifnextchar<%
{\tm \bk@op}%
{\rt.}%
}%
}%
}

\newcommand\Cases[1][0pt]{%
\def\Case@skipam{#1}%
\left\{\!\!\!\begin{array}{ll}\Cases@continue%
}

\def\Cases@continue#1#2{\@ifnextchar\bgroup%
{#1 &\txt{#2}\\[\Case@skipam] \Cases@continue}%
{#1 &\txt{#2}\end{array}\Cases@end}%
}

\def\Cases@end{\@ifnextchar[%
{\Cases@end@postsq}%
{\right.}
}

\def\Cases@end@postsq[#1]{\ifthenelse{\equal{\}}{#1}}
{\!\!\right\}}%
{\right.}
}

\newcommand{\MyMakeTheoMacros}[3]{
\expandafter\newcommand\csname\expandafter\@gobble\string#2NostarNoname@DGaux\endcsname[2][]
{\ifthenelse{\equal{}{##1}}%
{\begin{#1}~##2 \end{#1}}%
{\begin{#1}\label{##1}~##2\end{#1}}%
}
\expandafter\newcommand\csname\expandafter\@gobble\string#2StarNoname@DGaux\endcsname[1]
{\begin{#1*}~##1 \end{#1*}}
\newcommand#2{\expandafter\@ifstar%
\expandafter{\csname\expandafter\@gobble\string#2StarNoname@DGaux\endcsname}%
{\csname\expandafter\@gobble\string#2NostarNoname@DGaux\endcsname}%
}

\expandafter\newcommand\csname\expandafter\@gobble\string#2NostarName@DGaux\endcsname[3][]
{\ifthenelse{\equal{}{##1}}%
{\begin{#1}[\e{##2}]~##3 \end{#1}}%
{\begin{#1}[\e{##2}]\label{##1}~##3\end{#1}}%
}
\expandafter\newcommand\csname\expandafter\@gobble\string#2StarName@DGaux\endcsname[2]
{\begin{#1*}[\e{##1}]~##2 \end{#1*}}
\newcommand#3{\expandafter\@ifstar%
\expandafter{\csname\expandafter\@gobble\string#2StarName@DGaux\endcsname}
{\csname\expandafter\@gobble\string#2NostarName@DGaux\endcsname}%
}
}

\DeclareRobustCommand\dgref[2]{\@ifnextchar[%
{#2\dgref@presq}%
{#1\dgref@nopresq}%
}

\def\dgref@presq[#1]{\@ifnextchar[%
{\ref{#1}, \dgref@presq}
{\ref{#1} and~\dgref@nopresq}
}

\def\dgref@nopresq#1{\ref{#1}}

\DeclareRobustCommand\bref{\@ifnextchar[%
{\bref@presq}%
{\bref@nopresq}%
}

\def\bref@presq[#1]{\@ifnextchar[%
{(\ref{#1}), \bref@presq}
{(\ref{#1}) and~\bref@nopresq}
}

\def\bref@nopresq#1{(\ref{#1})}

\makeatother

\newcommand{\NewOp}[2]{
\newcommand{#1}[2][]{\ensuremath{\operatorname{#2}##1\ifthenelse{\equal{}{##2}}{}{\lf(##2\rt)}}}
}

\NewOp{\dom}{dom}

\newcommand{\NewId}[2]{
\newcommand{#1}[1][]{\ensuremath{\ifthenelse{\equal{}{##1}}{\operatorname{\mathit{#2}}}{\operatorname{\mathit{#2##1}}}}}
}

\newcommand{\NewIdR}[2]{
\newcommand{#1}[1][]{\ensuremath{\ifthenelse{\equal{}{##1}}{\operatorname{\mathrm{#2}}}{\operatorname{\mathrm{#2##1}}}}}
}

\newcommand{\NewIdSm}[2]{\NewId{#1}%
{\mathchoice{#2}{#2}{\scriptscriptstyle{#2}}{\scalebox{0.75}{$\scriptscriptstyle{#2}$}}}%
}

\NewIdSm{\tp}{\top}
\NewIdSm{\bt}{\bot}

\newcommand{\NewM}[2]{
\newcommand{#1}{\ensuremath{#2}}
}

\NewM{\OI}{\set{0,1}}
\NewM{\II}{\set{-1,1}}
\NewM{\TF}{\set{\tp,\bt}}

\newcommand{\MyMakeRefMacros}[3]{\newcommand{#1}{\dgref{#2}{#3}}}

\ifndef{theorem}{}
\ifndef{theorem*}{\newtheorem*{theorem*}{Theorem}}

\MyMakeTheoMacros{theorem}{\theo}{\ntheo}

\MyMakeRefMacros{\theoref}{Theorem~}{Theorems~}

\ifndef{definition}{}
\ifndef{definition*}{\newtheorem*{definition*}{Definition}}

\MyMakeTheoMacros{definition}{\defi}{\ndefi}

\MyMakeRefMacros{\defiref}{Definition~}{Definitions~}

\ifndef{claim}{}
\ifndef{claim*}{\newtheorem*{claim*}{Claim}}

\MyMakeTheoMacros{claim}{\clm}{\nclm}

\MyMakeRefMacros{\clmref}{Claim~}{Claims~}

\ifndef{lemma}{}
\ifndef{lemma*}{\newtheorem*{lemma*}{Lemma}}

\MyMakeTheoMacros{lemma}{\lem}{\nlem}

\MyMakeRefMacros{\lemref}{Lemma~}{Lemmas~}

\ifndef{corollary}{}
\ifndef{corollary*}{\newtheorem*{corollary*}{Corollary}}

\MyMakeTheoMacros{corollary}{\crl}{\ncrl}

\MyMakeRefMacros{\crlref}{Corollary~}{Corollaries~}

\ifndef{fact}{}
\ifndef{fact*}{\newtheorem*{fact*}{Fact}}

\MyMakeTheoMacros{fact}{\fct}{\nfct}

\MyMakeRefMacros{\fctref}{Fact~}{Facts~}

\ifndef{question}{}
\ifndef{question*}{\newtheorem*{question*}{Question}}

\MyMakeTheoMacros{question}{\quest}{\nquest}

\MyMakeRefMacros{\questref}{Question~}{Questions~}

\ifndef{construction}{}
\ifndef{construction*}{\newtheorem*{construction*}{Construction}}

\MyMakeTheoMacros{construction}{\constr}{\nconstr}

\MyMakeRefMacros{\constrref}{Construction~}{Constructions~}

\renewcommand{\qedsymbol}{$\blacksquare$}

\DeclareRobustCommand{\prflabel}[1]{%
\ifstrempty{#1}%
{\renewcommand{\qedsymbol}{$\blacksquare$}}%
{\renewcommand{\qedsymbol}{$\blacksquare_{\mbox{\it{\scriptsize{#1}}}}$}}%
}
\newcommand{\prfstart}[1][]{\ifthenelse{\equal{}{#1}}%
{\begin{proof}[\underline{\proofname}]\prflabel{}}%
{\begin{proof}[\underline{\proofname\ #1}]\prflabel{#1}}%
}

\newcommand{\prfend}[1][*]{%
\ifthenelse{\equal{*}{#1}}{}{\prflabel{#1}}%
\end{proof}%
}

\newcommand{\abstart}{\begin{abstract}}
\newcommand{\abend}{\end{abstract}}

\newcommand{\itstart}[1][]{%
\ifthenelse{\equal{}{#1}}%
{\begin{itemize}}%
{\begin{itemize}[#1]}%
}
\newcommand{\itend}{\end{itemize}}

\newcommand{\NewSectLike}[3]{
\NewcommandThreeArgsTwoOpt{#1}
{\ifthenelse{\equal{*}{##2}}{#2*}{#2{##2}}}
{#2{##2}\label{##1}}
{#2[##3]{##2}#3{##3}}
{#2[##3]{##2}\label{##1}#3{##3}}
}

\NewSectLike{\sect}{\section}{\sectionmark}

\NewSectLike{\ssect}{\subsection}{\subsectionmark}

\NewSectLike{\sssect}{\subsubsection}{\subsubsectionmark}

\MyMakeRefMacros{\sref}{Section~}{Sections~}
\MyMakeRefMacros{\ssref}{Subsection~}{Subsections~}

\newcommand{\toc}[3][]{%
\ifthenelse{\equal{chapter}{#2}}{\newpage}{}%
\ifthenelse{\equal{part}{#2}}{\newpage}{}%
\ifthenelse{\equal{}{#1}}%
{\phantomsection\addcontentsline{toc}{#2}{#3}}%
{\phantomsection\refstepcounter{#2}\label{#1}\addcontentsline{toc}{#2}{#3}}%
}

\newcommand{\toct}[1]{\toc{section}{#1}}

\makeatletter
\newif\ifdg@bibtoc
\newcommand{\bibtoc}{%
\ifdg@bibtoc\else
\apptocmd{\thebibliography}{\toct{\refname}}%
{\dg@bibtoctrue}%
{\PackageError{dgB}{Cannot add the bibliography TOC entry}%
{Call \string\bibtoc\space after bibliography-related packages.}}%
\fi
}
\makeatother

\newcommand{\bibstart}[1][99]{\bibtoc\begin{thebibliography}{#1}}
\newcommand{\bibend}{\end{thebibliography}}

\MyMakeRefMacros{\fnref}{Footnote~}{Footnotes~}

\newcommand{\m}[2][]{\MyChangeMathMargins%
\ifthenelse{\boolean{delim_grow}}{\delimiterfactor=1001}{}%
\ifthenelse{\equal{}{#1}}%
{\begin{align*} #2 \end{align*}}%
{\ifthenelse{\equal{P}{#1}}%
{\allowdisplaybreaks\begin{align*} #2%
\end{align*}\interdisplaylinepenalty=10000}%
{\ifthenelse{\equal{I}{#1}}%
{$#2$}%
{\begin{align}\begin{split}\label{#1} #2 \end{split}\end{align}}%
}%
}%
}

\makeatletter

\let\dgampersand\&

\DeclareRobustCommand\&{%
\new@ifnextchar[%
{\dgsep@reposit}%
{\dgampersand}%
}

\def\dgsep@reposit[#1]{\hspace{#1}&\hspace{-#1}}

\makeatother

\mathchardef\breakingcomma\mathcode`\,
{\catcode`,=\active
\gdef,{\breakingcomma\discretionary{}{}{}}
}

\newcommand{\lf}{\mathopen{}\mathclose\bgroup\left}
\newcommand{\rt}{\aftergroup\egroup\right}

\providecommand{\middle}{\big}
\newcommand{\md}{\middle}

\newcommand{\ud}{\vphantom{|_1^1}}

\makeatletter

\newcommand{\NewPar}[4][*]{
\ifthenelse{\equal{}{#1}}
{\def\dg@parfactory{\NewcommandThreeArgsTwoOpt[]}}
{\def\dg@parfactory{\NewcommandThreeArgsTwoOpt}}
\dg@parfactory{#2}
{\lf#3{##2}\rt#4}
{#3{##2}#4}
{\lf#3{##2}\rt#4_{\!##3}}
{#3{##2}#4_{\!##3}}
}
\makeatother

\NewPar{\lrp}{(}{)}
\NewPar{\lra}{\langle}{\rangle}
\NewPar[]{\sz}{|}{|}
\NewPar[]{\norm}{\|}{\|}
\NewPar[]{\ceil}{\lceil}{\rceil}
\NewPar[]{\floor}{\lfloor}{\rfloor}
\NewPar{\round}{\lfloor}{\rceil}

\NewcommandThreeArgsTwoOpt[]{\set}
{\lf\{#2\rt\}}
{}
{\lf\{#2\md|\ud#3\rt\}}
{\lf\{#2\ud#1\ud#3\rt\}}

\definecolor{Red}{rgb}{1,0,0}
\definecolor{DarkRed}{rgb}{0.54,0,0}
\definecolor{DarkGreen}{rgb}{0, 0.52, 0.05}
\definecolor{LightBlue}{rgb}{0.122,0.016,0.855}

\definecolor{dgProbar}{rgb}{0.3,0.3,0.3}%
\definecolor{dgPreBack}{RGB}{209,211,214}%

\DeclareTextFontCommand{\bemph}{\bfseries}
\DeclareTextFontCommand{\ibemph}{\bfseries\em}
\DeclareTextFontCommand{\ttl}{\ttfamily}

\ifthenelse{\boolean{are_slides}}{

\newcommand{\e}{\ei}
\newcommand<>{\ei}[1]{\IfMathMode%
{\only#2{\textcolor{BrickRed}}{#1}}%
{\emph{\only#2{\textcolor{BrickRed}}{#1}}}%
}
\newcommand<>{\eb}[1]{\IfMathMode%
{\only#2{\textcolor{BrickRed}}{\pmb{#1}}}%
{\bemph{\only#2{\textcolor{BrickRed}}{#1}}}%
}
\newcommand<>{\ebi}[1]{\IfMathMode%
{\only#2{\textcolor{BrickRed}}{\pmb{#1}}}%
{\ibemph{\only#2{\textcolor{BrickRed}}{#1}}}%
}
\newcommand<>{\eu}[1]{\IfMathMode%
{\only#2{\textcolor{BrickRed}}{\txt{\ul{#1}}}}%
{\only#2{\textcolor{BrickRed}}{\ul{#1}}}%
}
\newcommand<>{\eui}[1]{\IfMathMode%
{\only#2{\textcolor{BrickRed}}{\txt{\ul{#1}}}}%
{\emph{\only#2{\textcolor{BrickRed}}{\ul{#1}}}}%
}
\newcommand<>{\ebu}[1]{\IfMathMode%
{\only#2{\textcolor{BrickRed}}{\underline{\pmb{#1}}}}%
{\bemph{\only#2{\textcolor{BrickRed}}{\ul{#1}}}}%
}
\newcommand<>{\ebui}[1]{\IfMathMode%
{\only#2{\textcolor{BrickRed}}{\underline{#1}}}%
{\ibemph{\only#2{\textcolor{BrickRed}}{\ul{#1}}}}%
}
}{%
\newcommand{\e}{\emph}
\newcommand{\eb}{\bemph}
\newcommand{\ebi}{\ibemph}
\newcommand{\eu}[1]{\ttl{\ul{#1}}}
\newcommand{\eui}[1]{\emph{\ul{#1}}}
\newcommand{\ebu}[1]{\bemph{\ul{#1}}}
\newcommand{\ebui}[1]{\ibemph{\ul{#1}}}
}%

\DeclareRobustCommand{\ruin}[1]{\par\smallskip\noindent\underline{\emph{#1}}}

\newcommand{\txt}[1]{\textrm{#1}}%

\newcommand{\tm}{\cdot}

\NewcommandFourArgsTwoOptDefa{\overlay}
{0mu}{1}
{
\begingroup
\mathchoice{%
\bgroup
\scalebox{#4}{\ooalign{$\displaystyle#2\mkern#1$\cr
\hidewidth{$\displaystyle\mkern#1#3$}\hidewidth}}
\egroup
}{%
\bgroup
\scalebox{#4}{\ooalign{$\textstyle#2\mkern#1$\cr
\hidewidth{$\textstyle\mkern#1#3$}\hidewidth}}
\egroup
}{%
\bgroup
\scalebox{#4}{\ooalign{$\scriptstyle#2\mkern#1$\cr
\hidewidth{$\scriptstyle\mkern#1#3$}\hidewidth}}
\egroup
}{%
\bgroup
\scalebox{#4}{\ooalign{$\scriptscriptstyle#2\mkern#1$\cr
\hidewidth{$\scriptscriptstyle\mkern#1#3$}\hidewidth}}
\egroup
}
\endgroup
}

\makeatletter
\DeclareRobustCommand{\qto}{%
\begingroup
\m@th
\def\dg@quantumwave{\mathchoice
{\raisebox{.9ex}{$\displaystyle\sim$}}
{\raisebox{.9ex}{$\textstyle\sim$}}
{\raisebox{.9\fontdimen5\scriptfont2}{$\scriptstyle\sim$}}
{\raisebox{.9\fontdimen5\scriptscriptfont2}{$\scriptscriptstyle\sim$}}}%
\mathrel{\vphantom{\dg@quantumwave}\overlay{\rightarrow}
{\mkern-2.25mu\dg@quantumwave\mkern2.25mu}}%
\endgroup
}
\makeatother
\newcommand{\unin}{\mathrel{\overlay[1.25mu]{\subset}{\sim}}}

\NewcommandTwoArgsLastOpt{\GF}
{{\mathcal{G\!F}_{#1}}}
{{\mathcal{G\!F}_{#1}^{#2}}}

\def\?{\mskip 0.5\thinmuskip}%

\makeatletter
\ifthenelse{\boolean{are_slides}}{

\NewcommandTwoArgsLastOpt{\ssframe}
{\only<1>{\ssect{#1}}\frametitle{#1}}
{\only<1>{\ssect{#1}[#2]}\frametitle{#1}}

\newdimen\progressbar@width
\newcount\progressbar@mypage
\newcount\progressbar@topage

\setbeamertemplate{footline}{
\color{dgProbar}

}

\setbeamercolor{background canvas}{bg=dgPreBack}
}{}%
\makeatother

\numberwithin{theorem}{section}

\renewenvironment{abstract}{%
\small
\begin{center}
{\bfseries\abstractname\vspace{-.5em}\vspace{0pt}}
\end{center}
\list{}{\leftmargin1em\rightmargin\leftmargin}
\item\relax
}{\endlist}

\makeatletter
\let\dg@articlemaketitle\maketitle
\renewcommand{\maketitle}{{\let\rlap\hbox\dg@articlemaketitle}}
\makeatother

\hypersetup{
pdftitle={On the quantum communication complexity of total functions},
pdfauthor={Dmytro Gavinsky}}

\NewM{\Shape}{\mathsf{Shape}}
\NewM{\CheatShape}{\mathsf{Cheat\text{-}Shape}}

\title{On the quantum communication complexity of total functions}
\author{Dmytro Gavinsky\thanks{In 2022 the author changed his first name, as explained on his Internet page.}%
\thanks{This work was developed in collaboration with {\fontencoding{T1}\fontfamily{lmr}\selectfont\textsc{Anna}} -- Anonymous Neural-Network Assistant -- whom the author sincerely thanks.}\\[0.4ex]
{\small Institute of Mathematics of the Czech Academy of Sciences,
\v{Z}itn\'a 25, Praha~1, Czechia}\\[-0.1ex]
{\small IonQ Quantum, Inc., 4505 Campus Drive,
College Park, MD 20740, USA}}
\date{}

\begin{document}
\maketitle
\thispagestyle{empty}

\abstart
\noindent We present a total function with a poly-logarithmic two-message quantum protocol, whereas every randomised protocol, even with arbitrarily many rounds, requires polynomial communication.
\abend

\sect{Introduction}

Communication complexity is among the most productive computational settings for comparing quantum mechanics with classical information processing: it is one of the strongest regimes where unconditional lower bounds can be established, while remaining expressive enough to accommodate rather involved protocols.
This work concerns the most basic scenario -- two players solving a single problem.
Here is a brief informal introduction to the relevant concepts.

\itstart
\item In the model of bipartite communication there are two players, Alice and Bob, who each receive one part of the input: Alice gets $x$ and Bob gets $y$.
Their goal is to use the allowed type of communication in order to compute an answer that is correct with respect to the received pair $(x,y)$.
\item The three main bipartite layouts are \emph{two-way} (interactive) communication, \emph{one-way} communication, and \emph{simultaneous message passing} (SMP).
In the first case the players can exchange messages interactively before answering; in the second only Alice can send a message to Bob, who then answers; in the third both players send one message each to a referee, who then answers.
\item The three main types of problems are \emph{total functions}, \emph{partial functions}, and \emph{relations}: in the first case there is exactly one correct answer for each pair of input values, and the set of those pairs is the full product of the players' input sets; the second is similar, but the set of admissible pairs can be arbitrary; in the third, multiple correct answers to the same input pair are allowed.
\item An \emph{efficient} solution is a protocol in which the players use at most poly-logarithmic (in the input length) communication and produce a correct answer with high confidence.
\item Communication models can be strengthened by \emph{shared randomness} or by \emph{shared entanglement}, resources distributed before the protocol starts.
\itend

Quantum communication complexity has been an active area of research for several decades.
Among its many results, the most relevant to this work are the following structural separations.

\itstart
\item In 1998 a partial function was demonstrated \cite{BCW98} for which, in the zero-error regime, quantum protocols had an exponential advantage over the classical ones (both one-way and interactive).
\item In 1999 a partial function was demonstrated \cite{Raz99} that had an efficient quantum two-way protocol, but no efficient classical two-way protocol.
\item In 2001 a total function was demonstrated \cite{BCWdW01} that had an efficient quantum simulta\-neous-mes\-sages protocol without shared randomness, but no efficient classical simulta\-neous-mes\-sages protocol without shared randomness.
\item In 2004 a relation was demonstrated \cite{BYJK04} that had an efficient quantum simulta\-neous-mes\-sages protocol without shared randomness, but no efficient classical one-way protocol.
\item In 2006 a partial function was demonstrated \cite{GKKRW08} that had an efficient quantum one-way protocol, but no efficient classical one-way protocol.
\item In 2007 a relation was demonstrated \cite{Gavinsky2008} with an efficient quantum one-way protocol, but no efficient classical two-way protocol.
\item In 2010 a partial function was demonstrated \cite{KlartagRegev2011} with an efficient quantum one-way protocol, but no efficient classical two-way protocol.
\item In 2016 a partial function was demonstrated \cite{G15} with an efficient quantum simulta\-neous-mes\-sages protocol with shared entanglement, but no efficient classical two-way protocol.
\item In 2017 a partial function was demonstrated \cite{Gavinsky2019} with an efficient quantum simulta\-neous-mes\-sages protocol without shared randomness, but no efficient classical simulta\-neous-mes\-sages protocol, even with shared randomness.
\item In 2019 a relation was demonstrated \cite{Gavinsky2020bare} with an efficient quantum simulta\-neous-mes\-sages protocol without shared resources, but no efficient classical two-way protocol.
\itend

A separate line of work studies polynomial separations between randomised and quantum complexity for total Boolean functions, in both query and communication models \cite{AnshuEtAlLookup,BansalSinha,SherstovStorozhenkoWu,HasegawaLeGall}.

\ssect{Our result}

We construct an explicit family $\CheatShape_n$ of total Boolean functions on $N_n$ input bits per party.
It has a two-message protocol of cost in $O(\log^3N_n)$, where the first message is quantum and the second is classical.
Every interactive randomised protocol, on the other hand, has cost in $\Omega(N_n^{1/6}/\log^{4/3}N_n)$.

Our construction is conceptually indebted to the communication cheat-sheet framework of Anshu, Belovs, Ben-David, G\"o\"os, Jain, Kothari, Lee, and Santha \cite{AnshuEtAlLookup}, and to the multiplication-gate fully linear PCP of Boneh, Boyle, Corrigan-Gibbs, Gilboa, and Ishai \cite{BonehEtAlFLPCP}.
The underlying communication problem was partially inspired by \cite{G15}.
We take the needed technical ingredients of \cite{BonehEtAlFLPCP} from the author-posted extended manuscript; for transparency and verifiability, we re-derive those ingredients within our proofs.

\sect{Models and conventions}

Throughout, $n$ is a positive integer.
By default, $\log$ denotes the base-two logarithm and all entropies are measured in bits.
All arithmetic on coordinate indices in $\mathbb Z_n$ is modulo $n$.
For a string $u\in\{0,1\}^n$, coordinates are indexed by $\mathbb Z_n$ and $\sz{u}$ denotes its Hamming weight.
The domain of a partial function $f$ is $\dom{} f=\{(x,y):f(x,y)\ne *\}$.
For a positive integer $m$, write $[m]=\{1,\ldots,m\}$.
For a statement $E$, write $[E]\in\{0,1\}$ for its indicator.
For a probability distribution $P$ on a finite set, let $H(P)=-\sum_uP(u)\log P(u)$, with $0\log0=0$.
For distributions $P,Q$ on the same finite set, their total variation distance is $\Delta(P,Q)=\frac12\sum_u|P(u)-Q(u)|$.
We write $X\unin S$ when $X$ is uniform on the finite set $S$.

A protocol computes a partial function with error $\varepsilon$ if its output is correct with probability at least $1-\varepsilon$ on every input in the domain.
The probability is over all random coins and, for a quantum protocol, all measurement outcomes.
The cost of a protocol is the maximum, over inputs in the domain and possible executions, of the total number of communicated bits and qubits.
The quantity $R_\varepsilon(f)$ is the minimum cost of a public-coin randomised protocol with arbitrarily many rounds and error at most $\varepsilon$.
The quantity $Q_\varepsilon(f)$ is defined analogously for quantum communication starting from a fixed product state.
Message patterns are denoted by arrows:\ the symbol $\qto$ denotes a quantum message and $\rightarrow$ a classical message; thus $A\qto B\rightarrow A$ means that Alice sends a quantum message, Bob replies classically, and Alice outputs.
A rectangle in a two-party input space is a set $\mathcal A\times\mathcal B$, where $\mathcal A$ and $\mathcal B$ are sets of Alice's and Bob's inputs, respectively.

\sect{The function Shape}

For $i\in\mathbb Z_n$, let $\sigma_i$ be the cyclic shift defined by $(\sigma_i z)_j=z_{j-i}$ for every $j\in\mathbb Z_n$.
For $x=(x_1,x_2)$ and $y=(y_1,y_2)$, where all four strings have length $n$, put
\m{
d_i(x,y)=\sz{\sigma_i(x_1)\oplus x_2\oplus
\sigma_i(y_1)\oplus y_2}.
}
Equivalently, $d_i(x,y)$ is the Hamming distance between $\sigma_i(x_1)\oplus x_2$ and $\sigma_i(y_1)\oplus y_2$.
The shifted approximate equality problem is the partial function
\m{
\Shape_n(x,y)=
\begin{cases}
1&\text{if }\min_{i\in\mathbb Z_n} d_i(x,y)\le 2n/5,\\
0&\text{if }7n/15\le d_i(x,y)\le 8n/15
\text{ for every }i,\\
*&\text{otherwise.}
\end{cases}
}
Thus a $1$-input has at least one shift of distance at most $2n/5$, a $0$-input has every shifted distance in $[7n/15,8n/15]$, and all remaining inputs lie outside the promise.

\ssect{Quantum protocol for $\Shape$}[Quantum protocol for Shape]

\lem[lem:shape-quantum]
{
For every $0<\varepsilon<1/2$, the function $\Shape_n$ has an $A\qto B$ protocol of error at most $\varepsilon$ and cost in $O\bigl(\log n(\log n+\log(1/\varepsilon))\bigr)$.
}

\prfstart
\ruin{Shift-testing protocol}.
For $u\in\{0,1\}^n$, let
\m{
\bk{u}=\frac1{\sqrt n}
\sum_{j\in\mathbb Z_n}(-1)^{u_j}\bk{j}.
}
One copy of Alice's message consists of two registers, containing $\bk{x_1}$ and $\bk{x_2}$, and each register has $\lceil\log n\rceil$ qubits.
Bob multiplies the amplitude of $\bk{j}$ in the two registers by $(-1)^{y_{1,j}}$ and $(-1)^{y_{2,j}}$, respectively.
If $a=x_1\oplus y_1$ and $b=x_2\oplus y_2$, these phase operations transform the two registers into $\bk{a}$ and $\bk{b}$.

For a fixed shift $i$, Bob applies the unitary $\bk{j}\mapsto\bk{j+i}$ to $\bk{a}$ and performs a SWAP test against $\bk{b}$.
As $\sz{\sigma_i(a)\oplus b}=d_i(x,y)$, the inner product of the two registers is
\m{
\frac1n\sum_{j\in\mathbb Z_n}
(-1)^{a_{j-i}+b_j}=1-\frac{2d_i(x,y)}n
}
and the SWAP test accepts with probability
\m{
p_i=\frac{1+|1-2d_i(x,y)/n|^2}{2}.
}
If $d_i(x,y)\le2n/5$, then $1-2d_i(x,y)/n\ge1/5$, whereas $7n/15\le d_i(x,y)\le8n/15$ implies $|1-2d_i(x,y)/n|\le1/15$.
Consequently,
\m[eq:shape-swap-gap]{
d_i(x,y)\le2n/5&\quad\Longrightarrow\quad p_i\ge117/225,\\
7n/15\le d_i(x,y)\le8n/15
&\quad\Longrightarrow\quad p_i\le113/225.
}

Alice sends $t$ independent copies of the two-register state; Bob performs the described operation on each copy and accepts (answers ``$1$'') if at least a $115/225$ fraction of these $t$ tests accept.
Denote by $\Lambda_i$ the projector onto the corresponding subspace.
To succeed in distinguishing the two cases of \eqref{eq:shape-swap-gap} with error at most $\eta$,
\m{
t\in O(\log(1/\eta))
}
is sufficient.

\ruin{Combining all shifts}.
Write
\m{
\bk{\psi}=(\bk{a}\otimes\bk{b})^{\otimes t}
}
for the complete message state after Bob's phase operations.
Then
\m[eq:shape-amplified-shifts]{
d_i(x,y)\le2n/5
&\quad\Longrightarrow\quad\|\Lambda_i\psi\|^2\ge1-\eta,\\
7n/15\le d_i(x,y)\le8n/15
&\quad\Longrightarrow\quad\|\Lambda_i\psi\|^2\le\eta.
}
For
\m{
S=\sum_{i\in\mathbb Z_n}\Lambda_i,\qquad
\theta=\frac{\varepsilon}{4},\qquad
\eta=\frac{\varepsilon^2}{16n},
}
let $P$ be the projector onto the subspace spanned by the eigenvectors of $S$ with eigenvalues at least $\theta$ (note that $S\succeq 0$).
Bob measures the $t$ copies received from Alice with $\{P,I-P\}$ and outputs $1$ precisely on outcome $P$.
For $Q=I-P$, $QSQ\le\theta Q$, and $\Lambda_i\le S$ gives $Q\Lambda_iQ\le\theta Q$ and $\|Q\Lambda_i\|^2 \le\theta$.

On a $1$-input, choose any shift $i$ with $d_i(x,y)\le2n/5$.
Equation \eqref{eq:shape-amplified-shifts} gives $\|(I-\Lambda_i)\psi\|\le\sqrt\eta$.
Consequently,
\m{
\|Q\psi\|
\le\|Q\Lambda_i\psi\|+\|Q(I-\Lambda_i)\psi\|
\le\sqrt\theta+\sqrt\eta.
}
Since $\eta\le\varepsilon/16$, the rejection probability is at most $(\sqrt\theta+\sqrt\eta)^2 <\varepsilon$.

On a $0$-input, all $n$ projectors satisfy the second line of \eqref{eq:shape-amplified-shifts}.
Trivially, $P\le S/\theta$ and hence the acceptance probability is at most $\frac1\theta\sum_i\|\Lambda_i\psi\|^2 \le \frac{\varepsilon}4$, as required.
\prfend[\lemref{lem:shape-quantum}]

\sect{Overview of the argument}

Let
\m{
k=\lceil\log(2n+1)\rceil
}
and take $k$ instances of $\Shape_n$.
On a tuple satisfying all $k$ promises, their answers form an address $L\in\{0,1\}^k$, selecting one of $2^k$ drawers.
Each drawer is shared between Alice and Bob.
Its XOR contains a candidate output bit and a certificate asserting that the complete tuple of base inputs is legal.
The value of $\CheatShape_n$ is the candidate bit in the addressed drawer when its certificate is valid, and zero otherwise.
In particular, an off-promise tuple makes every certificate invalid, so the resulting Boolean function is total.

The legality certificate records the computation of the concrete promise verifier over a sufficiently large finite characteristic-two field $K$ -- in particular, when a XOR-decomposition of a field value is shared between the players, they need to add their shares in order to recover it.
Let $q$ be the number of multiplication gates in its arithmetic implementation, with gate $q$ carrying a final value that is zero precisely when the tuple is legal, and fix distinct points $\tau_1,\ldots,\tau_q\in K$.
A polynomial $p$ of degree at most $2q-2$ records the purported outputs of these gates at the corresponding points.
Together with the parties' inputs, these values determine the purported inputs to all multiplication gates; let $f_p$ and $g_p$ of degrees at most $q-1$ be their unique interpolants.
Validity requires the exact identity
\m{
p(T)\equiv f_p(T)g_p(T)\quad\text{in }K[T],
}
as well as $p(\tau_q)=0$.
For every fixed field point $t$, the values $p(t)$ and $p(\tau_q)$ are affine in the certificate, and the values $f_p(t)$ and $g_p(t)$ are affine in the parties' inputs (for the instances of $\Shape_n$) and the certificate (the latter provides, when needed for computing the input to a multiplication gate, the output values of earlier multiplication gates).
So, Bob can send additive shares that allow Alice to perform the test.
The point $t$ is sampled if and when the drawer to be tested -- containing the purported coefficients of $p$ in the XOR-decomposed form -- has been fixed by Bob, using his part of the input and Alice's quantum message.
Note that $p(t)$ is always computable using the purported $p$'s coefficients contained in the certificate from the fixed drawer; the values of $f_p(t)$ and $g_p(t)$, on the other hand, are computed directly (``semantically'') when $t\in\{\tau_1,\ldots,\tau_q\}$ and by interpolation otherwise.

For the quantum protocol, Alice sends to Bob quantum one-way messages for all $k$ instances of $\Shape_n$.
Bob solves the $k$ instances, thereby fixing the candidate address and the selected drawer, and then samples $t$.
His classical response contains the address, his share of the candidate output bit, the challenge point $t$ and his shares of the four field values.
The first message has $O(\log^3 n)$ qubits and the response has $O(\log n)$ bits.

For the classical lower bound, we compare two distributions, supported on $0$-inputs and $1$-inputs.
A rectangle of sufficiently large uniform mass cannot impose much restriction on most individual drawers.
This effectively reduces the analysis to that of a single instance of $\Shape_n$ -- the corresponding rectangle estimates in \cite{G15} give the lower bound $\Omega(\sqrt n/\log n)$ and \theoref{thm:main}.

\sect[sec:certificate-scheme]{Certifying legal Shape tuples}

Let
\m{
\mathbf x=(x^{(1)},\ldots,x^{(k)}),
\qquad
\mathbf y=(y^{(1)},\ldots,y^{(k)})
}
denote Alice's and Bob's $k$ base inputs, respectively.
Each $x^{(r)}$ and $y^{(r)}$ has $2n$ bits.
Write $\mathcal X_n=\mathcal Y_n=(\{0,1\}^{2n})^k$ for the two base input spaces.
Define
\m[eq:domain-definition]{
D_{n,k}(\mathbf x,\mathbf y)=1
\quad\Longleftrightarrow\quad
(x^{(r)},y^{(r)})\in\dom{\Shape_n}
\text{ for every }r\in[k].
}
We construct a deterministic predicate $V_{D_{n,k}}(\mathbf x,\mathbf y;c)$, whose certificate strings have length $\ell_n$, such that
\m[eq:certificate-existence]{
\exists c\in\{0,1\}^{\ell_n}\colon
V_{D_{n,k}}(\mathbf x,\mathbf y;c)=1
\quad\Longleftrightarrow\quad
D_{n,k}(\mathbf x,\mathbf y)=1.
}
Thus some certificate is accepted exactly when every base pair satisfies the $\Shape_n$ promise.

Writing $d_{r,i}=d_i(x^{(r)},y^{(r)})$, the predicate in \eqref{eq:domain-definition} is exactly
\m[eq:domain-circuit]{
D_{n,k}=\bigwedge_{r=1}^k\left(
\bigvee_{i\in\mathbb Z_n}[d_{r,i}\le\lfloor2n/5\rfloor]
\ \vee\!
\bigwedge_{i\in\mathbb Z_n}
[\lceil7n/15\rceil\le d_{r,i}\le\lfloor8n/15\rfloor]
\right).
}

Fix a fan-in-two Boolean circuit $C_{n,k}$ computing $1-D_{n,k}$.
For each $(r,i)$, wiring and XOR gates produce the residual bits, an $O(n)$-gate adder computes $d_{r,i}$, and $O(\log n)$ gates compare it with a fixed threshold.
The Boolean operations in \eqref{eq:domain-circuit} add $O(kn)$ gates, so we may fix $C_{n,k}$ with size
\m{
S_n\in O(kn^2).
}

\nlem[lem:certificate]{based on \cite{BonehEtAlFLPCP}}
{
There is an explicit deterministic predicate $V_{D_{n,k}}(\mathbf x,\mathbf y;c)$ satisfying \eqref{eq:certificate-existence}, with
\m{
\ell_n\in O(S_n\log S_n).
}
If $c=c_A\oplus c_B$, then, after the two certificate shares have been fixed, Bob can sample a field point and send $O(\log S_n)$ classical bits allowing Alice to test the certificate.
A valid certificate passes surely, whereas
\m[eq:test-soundness]{
\Pr[\text{Alice accepts an invalid certificate}]<\frac1{64}.
}
}

\prfstart
Every Boolean gate can be replaced by a constant-size circuit over the functionally complete basis $\{\wedge,\oplus,\neg\}$, so this replacement preserves the size bound up to a constant factor.
We work over a field of characteristic two, where the three gates act on Boolean values as $uv$, $u+v$, and $1+u$, respectively.
Ensure that the final output $1-D_{n,k}$ is produced by a multiplication gate, and add inert multiplication gates if necessary so that there are at least two multiplication gates.
Order the resulting $q\in O(S_n)$ multiplication gates so that every gate follows all multiplication gates on which it depends and let the output multiplication gate be the last one.

Between multiplication gates the circuit uses only additions and constants, so every wire entering multiplication gate $j$ is an affine function of $\mathbf x,\mathbf y$ and the outputs of the earlier multiplication gates.
Denote the two resulting affine forms by $\alpha_j$ and $\beta_j$.

Put
\m{
s_n=\left\lceil\log(64(2q-2)+1)\right\rceil,
\qquad K=\mathbb F_{2^{s_n}},
}
so $|K|=2^{s_n}\ge64(2q-2)+1>q$.
Fix distinct points $\tau_1,\ldots,\tau_q\in K$ and a binary basis of $K$.
A certificate is the binary coefficient vector of a polynomial $p\in K[T]$ of degree at most $2q-2$.
It therefore contains $2q-1$ field elements, and consequently
\m{
\ell_n=(2q-1)s_n\in O(S_n\log S_n).
}

Every certificate of the right length has a unique interpretation as $p$.
Regard $p(\tau_j)$ as the purported output of multiplication gate $j$, and put
\m{
a_j^p=\alpha_j(\mathbf x,\mathbf y,p(\tau_1),\ldots,p(\tau_{j-1})),
\quad
b_j^p=\beta_j(\mathbf x,\mathbf y,p(\tau_1),\ldots,p(\tau_{j-1})).
}
The $q$ distinct interpolation points determine unique polynomials $f_p,g_p\in K[T]$ of degree at most $q-1$ satisfying $f_p(\tau_j)=a_j^p$ and $g_p(\tau_j)=b_j^p$ for every $j\in[q]$.
Define
\m[eq:certificate-predicate]{
V_{D_{n,k}}(\mathbf x,\mathbf y;c)=1
\quad\Longleftrightarrow\quad
p\equiv f_pg_p\ \text{in }K[T]
\quad\text{and}\quad p(\tau_q)=0.
}

Suppose first that $D_{n,k}=1$ -- i.e., all $k$ Shape-promises are satisfied -- and let $a_j,b_j$ be the actual inputs to multiplication gate $j$.
Let $f$ and $g$ be the polynomials of degree at most $q-1$ interpolating $(a_j)_{j=1}^q$ and $(b_j)_{j=1}^q$ and take $p=fg$.
Then $\deg p\le2q-2$, and $p(\tau_j)=a_jb_j$ is the actual output of each multiplication gate $j$.
It follows that the values used to define $a_j^p$ and $b_j^p$ are the actual earlier gate outputs, so $a_j^p=a_j$ and $b_j^p=b_j$ for every $j$.
Uniqueness of interpolation gives $f_p=f$ and $g_p=g$, while $p(\tau_q)=1-D_{n,k}=0$.
Thus the certificate defined by $p$ satisfies \eqref{eq:certificate-predicate}.

For the converse, suppose that \eqref{eq:certificate-predicate} holds.
At $j=1$, the forms $\alpha_1,\beta_1$ depend only on the actual input, so evaluating $p\equiv f_pg_p$ at $\tau_1$ shows that $p(\tau_1)$ is the actual output of gate $1$.
Inductively, if $p(\tau_1),\ldots,p(\tau_{j-1})$ are the actual earlier multiplication outputs, then $a_j^p,b_j^p$ are the actual inputs to gate $j$, and
\m{
p(\tau_j)=f_p(\tau_j)g_p(\tau_j)=a_j^pb_j^p
}
is its actual output.
Consequently, $p(\tau_q)=1-D_{n,k}$.
The additional condition $p(\tau_q)=0$ therefore implies $D_{n,k}=1$, completing the proof of \eqref{eq:certificate-existence}.

\ruin{Distributed test}.
For fixed $t\in K$, the value $p(t)$ is linear in the certificate coefficients.
The gate-input forms are affine, so every $a_j^p,b_j^p$ is affine jointly in the parties' inputs and the certificate.
Interpolation and evaluation at a fixed point are linear maps; therefore $f_p(t)$, $g_p(t)$ and $p(\tau_q)$ are affine in the same variables.
In characteristic two, the XOR decomposition $c=c_A\oplus c_B$ is also an additive sharing of the field coefficients.
Accordingly, each of $p(t), f_p(t), g_p(t)$ and $p(\tau_q)$ decomposes as Alice's term depending only on $(\mathbf x,c_A)$ plus Bob's term depending only on $(\mathbf y,c_B)$.

Independently of the fixed certificate shares, Bob samples $t\unin K$ and sends it together with his four additive shares.
Alice adds her corresponding shares, verifies that $p(\tau_q)=0$ and tests whether $p(t)=f_p(t)g_p(t)$ -- a valid certificate satisfies both identities and therefore passes.
For an invalid certificate, the test rejects surely if $p(\tau_q)\ne0$.
Otherwise, $p-f_pg_p$ is a non-zero polynomial of degree at most $2q-2$, so it vanishes at the sampled $t$ with probability at most
\m{
\frac{2q-2}{|K|}<\frac1{64}.
}
This proves \eqref{eq:test-soundness}.
The response consists of five field elements and hence has $5s_n\in O(\log S_n)$ classical bits, as required.
\prfend[\lemref{lem:certificate}]

\sect{The function $\CheatShape$}[The function Cheat-Shape]

Let $\mathbf x$, $\mathbf y$, $\ell_n$ and $V_{D_{n,k}}(\mathbf x,\mathbf y;c)$ be as in \sref{sec:certificate-scheme} and define the $k$-bit \emph{address} as
\m{
L=L(\mathbf x,\mathbf y)=
\begin{cases}
\bigl(\Shape_n(x^{(1)},y^{(1)}),\ldots,
\Shape_n(x^{(k)},y^{(k)})\bigr)
&\text{if }D_{n,k}(\mathbf x,\mathbf y)=1;\\
0^k&\text{if }D_{n,k}(\mathbf x,\mathbf y)=0.
\end{cases}
}
Let
\m{
\Gamma_n=\{0,1\}^{1+\ell_n}
}
be the set of possible drawer contents.
In addition to $\mathbf x$ and $\mathbf y$, the players receive tables
\m{
U=(U_j)_{j\in\{0,1\}^k},
\qquad V=(V_j)_{j\in\{0,1\}^k},
\qquad U_j,V_j\in\Gamma_n,
}
so that Alice's input is $(\mathbf x,U)$ and Bob's is $(\mathbf y,V)$.
For every $j\in\{0,1\}^k$, the pair $(U_j,V_j)$ forms a \emph{drawer}, whose contents are defined by
\m{
U_j\oplus V_j=(o_j,c_j),
\qquad o_j\in\{0,1\},\quad c_j\in\{0,1\}^{\ell_n},
}
where $c_j$ is the certificate and $o_j$ is the candidate output value of drawer $j$.
Then
\m{
\CheatShape_n(\mathbf x,U;\mathbf y,V)
=o_L\wedge V_{D_{n,k}}(\mathbf x,\mathbf y;c_L),
}
so that the number of input bits held by either party is
\m[eq:input-length]{
N_n = 2nk+2^k(1+\ell_n) \in O(n^3\log^2n).
}

By the preceding argument, $\CheatShape_n$ is a total function -- in particular, its value is $0$ if any Shape-promise is violated.
For each legal tuple $(\mathbf x,\mathbf y)$, fix a valid certificate $c^*(\mathbf x,\mathbf y)$ and put
\m{
w^*(\mathbf x,\mathbf y)=(1,c^*(\mathbf x,\mathbf y))\in\Gamma_n,
}
so that $U_L\oplus V_L=w^*(\mathbf x,\mathbf y)$ makes the function value $1$ (regardless of the other drawers).

\ssect{Quantum protocol for $\CheatShape$}[Quantum protocol for Cheat-Shape]

\lem[lem:cheat-shape-quantum]
{
The function $\CheatShape_n$ has an $A\qto B\rightarrow A$ protocol of error less than $1/32$ and cost in $O(\log^3n)$.
}

\prfstart
\ruin{Protocol}.
The protocol has communication pattern $A\qto B\rightarrow A$.
Set $\varepsilon=1/(64k)$ and let Alice send in parallel $k$ quantum messages of \lemref{lem:shape-quantum} -- one message per base instance, for error probability at most $\varepsilon$ each.
Their total length is in $O(\log^3n)$.

Bob measures these messages and obtains $J=(J_1,\ldots,J_k)\in\{0,1\}^k$, thereby fixing the drawer to be tested.
Write $U_J=(o_J^A,c_A)$ and $V_J=(o_J^B,c_B)$, so $o_J=o_J^A\oplus o_J^B$ and $c_J=c_A\oplus c_B$.
The selected certificate $c_J$ encodes a polynomial $p$ as in \lemref{lem:certificate}.
Then he samples $t\unin K$ and responds to Alice with $O(\log n)$ classical bits
\m{
J;\quad
o_J^B;\quad
t;\quad
\text{his shares of }p(t),f_p(t),g_p(t),p(\tau_q).
}
Alice selects $U_J$, combines $o_J^A$ with $o_J^B$ to reconstruct $o_J$ and combines her locally computed field shares with Bob's shares to reconstruct the four tested values.
She applies the certificate test, outputs $o_J$ on acceptance and ``$0$'' on rejection.

\ruin{Legal tuples}.
Suppose first that $D_{n,k}=1$, so every base pair satisfies the promise and $L$ is the vector of their correct answers.
The union bound gives
\m{
\Pr[J\ne L]
\le\sum_{r=1}^k\Pr[J_r\ne L_r]
\le k\varepsilon=\frac1{64}.
}
Condition on $J=L$:\ if $c_L$ is valid, the certificate test accepts surely and Alice outputs $o_L$; otherwise, Alice errs only if the test falsely accepts, which happens with probability less than $1/64$ by \eqref{eq:test-soundness}.
The total error on a legal tuple is therefore less than $1/32$.

\ruin{Off-promise tuples}.
If $D_{n,k}=0$, then any certificate is invalid by \eqref{eq:certificate-existence}; \eqref{eq:test-soundness} implies that Alice errs with probability less than $1/64$.
\prfend[\lemref{lem:cheat-shape-quantum}]

\sect{Randomised lower bound for $\CheatShape$}[Randomised lower bound for Cheat-Shape]

Fix a sufficiently large $n$.
Let $\mu_0$ be uniform on one joint $\Shape_n$ input $(X_1,X_2,Y_1,Y_2)\in\{0,1\}^{4n}$.
To sample $\mu_1$, choose $i\unin\mathbb Z_n$, draw the same quadruple uniformly conditioned on $d_i(X,Y)=0$ (i.e., $[\sigma_i(X_1\oplus Y_1) = X_2\oplus Y_2]$) and independently flip every coordinate of $X_1$ with probability $3/8$.
Put
\m{
\nu_b=\mu_b(\,\cdot\mid\Shape_n=b),
\qquad
\nu_z=\bigotimes_{r=1}^k\nu_{z_r}
\quad
(b\in\{0,1\},\ z\in\{0,1\}^k).
}

\lem[lem:shape-switch]
{
There are a constant $C>1$ and positive $\delta_n\in2^{-\Omega(\sqrt n)}$ such that, for all sufficiently large $n$, every $z\in\{0,1\}^k$, every $f:\mathcal X_n\to[0,1]$ and every $g:\mathcal Y_n\to[0,1]$ satisfy
\m[eq:shape-switch]{
\mathbb E_{\mu_0^{\otimes k}}f(\mathbf X)g(\mathbf Y)
\le C^k\cdot\mathbb E_{\nu_z}f(\mathbf X)g(\mathbf Y)+\delta_n.
}
}

\prfstart
\ruin{One instance}.
Lemmas~2 and~3 of~\cite{G15}, used with parameter $1/184$, imply that every one-copy rectangle $Q$ satisfies
\m[eq:one-copy-switch]{
\mu_0(Q)\le46\mu_1(Q)+2^{-\sqrt{n/184}/29}.
}
The claim is immediate when $\mu_0(Q)=0$, so suppose otherwise.
Indeed, if $\mu_1(Q)>\mu_0(Q)/46$, there is nothing to prove.
Otherwise, Lemma~2 of~\cite{G15} gives average noisy-shift entropy at most $n-1/184$ on one uniform side distribution, and Lemma~3 bounds its min-entropy by $2n-\sqrt{n/184}/29$.
The density of that side, and hence $\mu_0(Q)$, is therefore at most $2^{-\sqrt{n/184}/29}$.
The same inequality holds, trivially, for products of local $[0,1]$-valued functions.

\ruin{Several instances}.
Replace $\mu_0$ by $\mu_1$ in each coordinate $r$ with $z_r=1$.
At each step, fix all other base pairs and apply \eqref{eq:one-copy-switch} to the remaining local functions.
Averaging over the fixed pairs and repeating gives
\m{
\mathbb E_{\mu_0^{\otimes k}}fg
\le46^{\sz z}\mathbb E_{\bigotimes_r\mu_{z_r}}fg
+2^{-\sqrt{n/184}/29}\sum_{h=0}^{k-1}46^h.
}

\ruin{Conditioning on the promises}.
Under $\mu_0$, each $d_i$ has the binomial distribution with parameters $n,1/2$.
Hoeffding's inequality and a union bound over the $n$ shifts show that all distances lie in $[7n/15,8n/15]$, except with probability at most $2n\exp(-2n/30^2)$.
Under $\mu_1$, the distinguished shift has the binomial distribution with parameters $n,3/8$.
It exceeds $2n/5$ with probability at most $\exp(-2n/40^2)$.
Consequently,
\m{
\Pr_{\mu_b}[\Shape_n\ne b]\in e^{-\Omega(n)}
\qquad(b\in\{0,1\}).
}
Since $k\in O(\log n)$, the probability under $\bigotimes_r\mu_{z_r}$ that $\Shape_n(X^{(r)},Y^{(r)})=z_r$ for all $r\in\{1,\ldots,k\}$ is $1-e^{-\Omega(n)}$.
Conditioning on this event gives $\nu_z$ and, since $0\le fg\le1$, changes the expectation by at most $e^{-\Omega(n)}$.
After multiplication by $46^k$, this contributes only $e^{-\Omega(n)}$ to the additive error.
Thus equation~\eqref{eq:shape-switch} follows with $C=46$ and some $\delta_n\in2^{-\Omega(\sqrt n)}$.
\prfend[\lemref{lem:shape-switch}]

Let $\Omega_n$ be uniform on the complete input space and define
\m{
\Lambda_{0,n}=\Omega_n(\,\cdot\mid \CheatShape_n=0).
}
This conditioning event has probability at least $1/2$: an illegal base tuple always gives zero, while for each fixed legal tuple the candidate bit $o_L$ is unbiased.
To sample $\Lambda_{1,n}$, choose $J\unin\{0,1\}^k$, draw $(\mathbf X,\mathbf Y)$ from $\nu_J$, and draw $U\unin\Gamma_n^{2^k}$ and $V_{-J}=(V_i)_{i\in\{0,1\}^k\setminus\{J\}}\unin\Gamma_n^{2^k-1}$ independently.
Set
\m{
V_J=U_J\oplus w^*(\mathbf X,\mathbf Y).
}
Under $\nu_J$, the base tuple is legal and has address $J$.
These $\Lambda_{0,n}$ and $\Lambda_{1,n}$ are supported on $\CheatShape_n^{-1}(0)$ and $\CheatShape_n^{-1}(1)$, respectively.

\lem[lem:asymmetric-rectangle]
{
There are $A_n\in n^{O(1)}$ and $\eta_n\in2^{-\Omega(\sqrt n)}$ such that every rectangle $R$ satisfies
\m[eq:asymmetric-rectangle]{
\Lambda_{0,n}(R)\le A_n\cdot\Lambda_{1,n}(R)+\eta_n.
}
}

Intuitively, a rectangle of large uniform mass cannot substantially restrict most individual drawers -- therefore, for a drawer whose two shares are nearly uniform, fixing their XOR to an accepting value preserves a constant fraction of the mass.

\prfstart[of \lemref{lem:asymmetric-rectangle}]
Write $R=\mathcal A\times\mathcal B$.
The claim is immediate if $\Omega_n(R)=0$, so put
\m{
D=\log\frac1{\Omega_n(R)}.
}
For base inputs $\mathbf x,\mathbf y$, define
\m{
a(\mathbf x)=\Pr_{U\unin\Gamma_n^{2^k}}[(\mathbf x,U)\in\mathcal A],
\qquad
b(\mathbf y)=\Pr_{V\unin\Gamma_n^{2^k}}[(\mathbf y,V)\in\mathcal B].
}
For each $j\in\{0,1\}^k$ and each $\mathbf x$ with $a(\mathbf x)>0$, let $P^A_{\mathbf x,j}$ be the distribution of $U_j$ conditional on $(\mathbf x,U)\in\mathcal A$, with $U$ originally uniform on $\Gamma_n^{2^k}$; define $P^B_{\mathbf y,j}$ analogously.
When $a(\mathbf x)=0$ or $b(\mathbf y)=0$, respectively, choose the corresponding distributions arbitrarily.
For a distribution $P$ on $\Gamma_n$, write $d(P)=\log|\Gamma_n|-H(P)$ to denote the corresponding entropy deficit.

\ruin{The total restriction on the drawers}.
Entropy subadditivity gives
\m{
\log|\mathcal A|
&=H(\mathbf X,U\mid R)\\
&\le H(\mathbf X\mid R)+\sum_{j\in\{0,1\}^k}H(U_j\mid\mathbf X,R)\\
&\le\log|\mathcal X_n|+2^k\log|\Gamma_n|
-\mathbb E_{\Omega_n(\cdot\mid R)}\sum_{j\in\{0,1\}^k}d(P^A_{\mathbf X,j}).
}
The analogous inequality for Bob, together with $\Omega_n(R)=|\mathcal A||\mathcal B|/(|\mathcal X_n||\mathcal Y_n||\Gamma_n|^{2^{k+1}})$, yields
\m{
\mathbb E_{\Omega_n(\cdot\mid R)}\sum_{j\in\{0,1\}^k}
\bigl(d(P^A_{\mathbf X,j})+d(P^B_{\mathbf Y,j})\bigr)\le D.
}

Put $\kappa=1/64$ and, for each $j\in\{0,1\}^k$, define
\m{
f_j(\mathbf x)=a(\mathbf x)[d(P^A_{\mathbf x,j})\le\kappa],
\qquad
g_j(\mathbf y)=b(\mathbf y)[d(P^B_{\mathbf y,j})\le\kappa].
}
Since the base marginal of $\Omega_n(\,\cdot\mid R)$ has density $a(\mathbf x)b(\mathbf y)/\Omega_n(R)$ relative to $\mu_0^{\otimes k}$, Markov's inequality and the union bound give
\m[eq:retained-mass]{
\mathbb E_{j\unin\{0,1\}^k}\mathbb E_{\mu_0^{\otimes k}}f_j(\mathbf X)g_j(\mathbf Y)
\ge\Omega_n(R)\left(1-\frac D{\kappa 2^k}\right).
}
If $D>\kappa 2^{k-1}$, then $\Omega_n(R)<2^{-\kappa 2^{k-1}}$; hence suppose $D\le\kappa 2^{k-1}$.

\ruin{Fixing an accepting drawer}.
Let $W$ be uniform on $\Gamma_n$.
For any distributions $P,Q$ on $\Gamma_n$ and any $w\in\Gamma_n$,
\m[eq:overlap]{
|\Gamma_n|\sum_uP(u)Q(u\oplus w)
\ge1-\Delta(P,W)-\Delta(Q,W).
}
Indeed, apply $rs\ge\min\{r,1\}+\min\{s,1\}-1$ with $r=|\Gamma_n|P(u)$ and $s=|\Gamma_n|Q(u\oplus w)$, and average over uniform $u$.
Pinsker's inequality gives
\m{
\Delta(P,W)\le\sqrt{\frac{\ln2}{2}\,d(P)}\le\sqrt{d(P)}.
}
Thus the right-hand side of \eqref{eq:overlap} is at least $1/2$ whenever $d(P),d(Q)\le\kappa$.

For any legal base input $(\mathbf x,\mathbf y)$ with address $j\in\{0,1\}^k$, counting allowed completions of the other drawers in the sampling of $\Lambda_{1,n}$ gives
\m{
\Pr_{\Lambda_{1,n}}[R\mid\mathbf X=\mathbf x,\mathbf Y=\mathbf y,J=j]
=|\Gamma_n|a(\mathbf x)b(\mathbf y)
\sum_uP^A_{\mathbf x,j}(u)
P^B_{\mathbf y,j}(u\oplus w^*(\mathbf x,\mathbf y))
\ge \frac{f_j(\mathbf x)g_j(\mathbf y)}2.
}
Consequently, \lemref{lem:shape-switch} and \eqref{eq:retained-mass} give
\m{
\Lambda_{1,n}(R)
&\ge\frac12\mathbb E_{j\unin\{0,1\}^k}\mathbb E_{\nu_j}f_j(\mathbf X)g_j(\mathbf Y)\\
&\ge\frac{C^{-k}}2
\left(\Omega_n(R)\left(1-\frac D{\kappa 2^k}\right)-\delta_n\right).
}
Since $D\le\kappa 2^{k-1}$, it follows that $\Omega_n(R)\le4C^k\Lambda_{1,n}(R)+2\delta_n$.
Moreover, $\Lambda_{0,n}(R)\le2\Omega_n(R)$, since $\Omega_n[\CheatShape_n=0]\ge1/2$.
Combining this with the case $D>\kappa 2^{k-1}$ proves \eqref{eq:asymmetric-rectangle} with
\m{
A_n=8C^k,
\qquad
\eta_n=4\delta_n+2^{1-\kappa 2^{k-1}}.
}
The asserted asymptotics follow from $2^k\in\Theta(n)$ and $k\in O(\log n)$.
\prfend[\lemref{lem:asymmetric-rectangle}]

\theo[thm:main]
{
\m{
R_{1/3}(\CheatShape_n)
\in\Omega\!\left(\frac{N_n^{1/6}}{\log^{4/3}N_n}\right).
}
}

\prfstart
Let $c$ be the cost of a public-coin randomised protocol for $\CheatShape_n$ with error at most $1/3$.
Amplify it by independent repetitions to error
\m{
\varepsilon=\frac1{4(A_n+1)},
}
where $A_n$ is from \lemref{lem:asymmetric-rectangle}.
This uses $O(\log n)$ repetitions because $A_n\in n^{O(1)}$.
Append the final output bit so that the output is fixed on every transcript rectangle.
The resulting cost satisfies
\m{
c'\in O((c+1)\log n).
}
For each fixed choice of the public coins, sum \eqref{eq:asymmetric-rectangle} over the at most $2^{c'}$ transcript rectangles on which this protocol outputs zero.
The rectangles are disjoint, so averaging the resulting inequality over the coins gives
\m{
1-\varepsilon\le A_n\varepsilon+2^{c'}\eta_n.
}
Here the left side uses correctness on $\Lambda_{0,n}$, and the first term on the right uses the error bound on $\Lambda_{1,n}$.
Consequently,
\m{
2^{c'}\eta_n\ge1-(A_n+1)\varepsilon=\frac34.
}
Since $\eta_n\in2^{-\Omega(\sqrt n)}$, it follows that $c'\in\Omega(\sqrt n)$, and hence
\m{
c+1\in\Omega\!\left(\frac{\sqrt n}{\log n}\right).
}
The right side tends to infinity, so the same asymptotic lower bound holds for $c$.
This proves
\m{
R_{1/3}(\CheatShape_n)\in\Omega\!\left(\frac{\sqrt n}{\log n}\right).
}

Finally, \eqref{eq:input-length} gives $N_n^{1/6}\in O(\sqrt n\,\log^{1/3}n)$ and the claimed bound.
\prfend[\theoref{thm:main}]

\sect*{Acknowledgements}

\toct{Acknowledgements}

I am very grateful to Ronald de~Wolf and Fran\c{c}ois Le~Gall for helpful and encouraging discussions.
Fran\c{c}ois' interest in this work is also reflected in his recent follow-up \cite{LeGall}, which improves the quantitative bounds.

\bibstart
\small
\setlength{\itemsep}{0pt}

\bibitem{AnshuEtAlLookup}
Anurag Anshu, Aleksandrs Belovs, Shalev Ben-David, Mika G\"o\"os, Rahul Jain, Robin Kothari, Troy Lee, and Miklos Santha.
\newblock Separations in communication complexity using cheat sheets and information complexity.
\newblock In \emph{Proceedings of the 57th IEEE Symposium on Foundations of Computer Science (FOCS)}, pages 555--564, 2016.
\newblock \href{https://doi.org/10.1109/FOCS.2016.66}{doi:10.1109/FOCS.2016.66}; full version \href{https://arxiv.org/abs/1605.01142}{arXiv:1605.01142}.

\bibitem{BansalSinha}
Nikhil Bansal and Makrand Sinha.
\newblock $k$-Forrelation optimally separates quantum and classical query complexity.
\newblock In \emph{Proceedings of the 53rd ACM Symposium on Theory of Computing (STOC)}, pages 1303--1316, 2021.
\newblock \href{https://doi.org/10.1145/3406325.3451040}{doi:10.1145/3406325.3451040}.

\bibitem{BYJK04}
Ziv Bar-Yossef, T.~S. Jayram, and Iordanis Kerenidis.
\newblock Exponential separation of quantum and classical one-way communication complexity.
\newblock \emph{SIAM Journal on Computing}, 38(1):366--384, 2008.

\bibitem{BonehEtAlFLPCP}
Dan Boneh, Elette Boyle, Henry Corrigan-Gibbs, Niv Gilboa, and Yuval Ishai.
\newblock Zero-knowledge proofs on secret-shared data via fully linear PCPs.
\newblock In \emph{Advances in Cryptology -- CRYPTO 2019, Part III}, pages 67--97, 2019.
\newblock \href{https://doi.org/10.1007/978-3-030-26954-8_3}{doi:10.1007/978-3-030-26954-8\_3}; full version \href{https://eprint.iacr.org/2019/188}{IACR ePrint 2019/188}.

\bibitem{BCW98}
Harry Buhrman, Richard Cleve, and Avi Wigderson.
\newblock Quantum vs. classical communication and computation.
\newblock In \emph{Proceedings of the 30th ACM Symposium on Theory of Computing (STOC)}, pages 63--68, 1998.

\bibitem{BCWdW01}
Harry Buhrman, Richard Cleve, John Watrous, and Ronald de~Wolf.
\newblock Quantum fingerprinting.
\newblock \emph{Physical Review Letters}, 87(16):167902, 2001.

\bibitem{Gavinsky2008}
Dmytro Gavinsky.
\newblock Classical interaction cannot replace a quantum message.
\newblock In \emph{Proceedings of the 40th ACM Symposium on Theory of Computing (STOC)}, pages 95--102, 2008.

\bibitem{GKKRW08}
Dmytro Gavinsky, Julia Kempe, Iordanis Kerenidis, Ran Raz, and Ronald de~Wolf.
\newblock Exponential separation for one-way quantum communication complexity, with applications to cryptography.
\newblock \emph{SIAM Journal on Computing}, 38(5):1695--1708, 2008.

\bibitem{Gavinsky2019}
Dmytro Gavinsky.
\newblock Quantum versus classical simultaneity in communication complexity.
\newblock \emph{IEEE Transactions on Information Theory}, 65(10):6466--6483, 2019.

\bibitem{G15}
Dmytro Gavinsky.
\newblock Entangled simultaneity versus classical interactivity in communication complexity.
\newblock \emph{IEEE Transactions on Information Theory}, 66(7):4641--4651, 2020.
\newblock \href{https://doi.org/10.1109/TIT.2020.2976074}{doi:10.1109/TIT.2020.2976074}.

\bibitem{Gavinsky2020bare}
Dmytro Gavinsky.
\newblock Bare quantum simultaneity versus classical interactivity in communication complexity.
\newblock \emph{IEEE Transactions on Information Theory}, 67(10):6583--6605, 2021.

\bibitem{HasegawaLeGall}
Atsuya Hasegawa and Fran\c{c}ois Le~Gall.
\newblock Constant-round quantum advantage in communication complexity for total functions.
\newblock \href{https://arxiv.org/abs/2608.19787}{arXiv:2608.19787}, 2026.

\bibitem{KlartagRegev2011}
Bo'az Klartag and Oded Regev.
\newblock Quantum one-way communication can be exponentially stronger than classical communication.
\newblock In \emph{Proceedings of the 43rd ACM Symposium on Theory of Computing (STOC)}, pages 31--40, 2011.

\bibitem{LeGall}
Fran\c{c}ois Le~Gall.
\newblock Improved separations between quantum and classical communication complexity of total functions.
\newblock \href{https://arxiv.org/abs/2609.16726}{arXiv:2609.16726}, 2026.

\bibitem{Raz99}
Ran Raz.
\newblock Exponential separation of quantum and classical communication complexity.
\newblock In \emph{Proceedings of the 31st ACM Symposium on Theory of Computing (STOC)}, pages 358--367, 1999.

\bibitem{SherstovStorozhenkoWu}
Alexander~A. Sherstov, Andrey~A. Storozhenko, and Pei Wu.
\newblock An optimal separation of randomized and quantum query complexity.
\newblock \emph{SIAM Journal on Computing}, 52(2):525--567, 2023.
\newblock \href{https://doi.org/10.1137/22M1468943}{doi:10.1137/22M1468943}.

\bibend
\end{document}